\documentclass[aps,prl,reprint]{revtex4-1}
\usepackage{amsmath}
\usepackage[utf8]{inputenc}
\usepackage{amsmath,amssymb,mathtools}
\usepackage{graphicx}
\usepackage[hidelinks, pageanchor=false]{hyperref}
\usepackage{xcolor}
\usepackage{multirow}
\usepackage{textgreek}
\usepackage{babel}
\usepackage{xspace}
\usepackage{enumerate}
\usepackage{caption}
\usepackage{ragged2e}

\usepackage{float} 
\usepackage{comment}
\usepackage{orcidlink}
\usepackage[
  a4paper,
  twoside=false,
  left=1.7cm,
  right=1.7cm,
  top=2.5cm,
  bottom=3cm
]{geometry}

\renewcommand{\thefigure}{\arabic{figure}}  
\allowdisplaybreaks

\def\beq{\begin{equation}}
\def\eeq{\end{equation}}
\def\bea{\begin{align}}
\def\eea{\end{align}}

\def\Eq#1{Eq.~(\ref{#1})}
\def\ln#1{\mathrm{log}\left(#1\right)}

\def\ra{\rangle}

\def\ket#1{|#1\ra}

\footnotetext{These authors contributed equally to this work.}
\footnotetext{jormard@ific.uv.es}
\begin{document} 
\author{Jorge J. Mart\'{\i}nez de Lejarza\orcidlink{0000-0002-3866-3825}$^{1,\dagger, *}$}
\author{Hsin-Yu Wu\orcidlink{0000-0001-8201-4286}$^{2,3,\dagger}$}
\author{{Oleksandr Kyriienko\orcidlink{0000-0002-6259-6570}}$^{4}$}
\author{Germán Rodrigo\orcidlink{0000-0003-0451-0529}$^1$}
\author{Michele Grossi\orcidlink{0000-0003-1718-1314}$^5$}
\affiliation{%
 $^1$Instituto de F\'{\i}sica Corpuscular, Universitat de Val\`encia - Consejo Superior de Investigaciones Cient\'{\i}ficas, Parc Cient\'{\i}fic, E-46980 Paterna, Valencia, Spain}
\affiliation{$^2$Department of Physics and Astronomy, University of Exeter, Stocker Road, Exeter EX4 4QL, United Kingdom}
\affiliation{$^3$Pasqal, 7 rue Léonard de Vinci, 91300, Massy, France}
\affiliation{$^4$School of Mathematical and Physical Sciences, University of Sheffield, Sheffield S10 2TN, United Kingdom}
\affiliation{$^5$European Organization for Nuclear Research (CERN), 1211 Geneva, Switzerland}

\title{Quantum Chebyshev Probabilistic Models for Fragmentation Functions}

\date{\today}

\begin{abstract}
Quantum generative modeling is emerging as a powerful tool for advancing data analysis in high-energy physics, where complex multivariate distributions are common. However, efficiently learning and sampling these distributions remains challenging. We propose a quantum protocol for a bivariate probabilistic model based on shifted Chebyshev polynomials, trained as a circuit-based representation of two correlated variables, with sampling performed via quantum Chebyshev transforms. As a key application we study fragmentation functions (FFs) of charged pions and kaons from single-inclusive hadron production in electron-positron annihilation. We learn the joint distribution of momentum fraction $z$ and energy scale $Q$, and infer their correlations from the entanglement structure. Building on the generalization capabilities of the quantum model and extended register architecture, we perform fine-grid multivariate sampling for FF dataset augmentation. Our results highlight the growing potential of quantum generative modeling to advance data analysis and scientific discovery in high-energy physics.

\end{abstract}

\maketitle

\section*{Introduction} The ability to analyze scientific data relies on designing advanced machine learning tools, inferring correlations, and performing generative modeling by sampling synthetic distributions \cite{Wang2023}. Multivariate differential distributions represent a major interest in high-energy physics (HEP), specifically in the context of Quantum Chromodynamics (QCD) and studies of scattering processes~\cite{Guest2018}. Quantum computing shows promise for accelerating calculations of physical properties and is believed to be vital for large-scale QCD simulations in the future~\cite{Delgado:2022tpc, DiMeglio:2023nsa, Rodrigo:2024say}. Moreover, Quantum Machine Learning (QML) methods become increasingly more developed \cite{Biamonte2017,Cerezo2021rev,du2025QMLtutorial}. Here, HEP data analysis is a suitable example of learning processes that are inherently quantum. For instance, to date QML was applied to jet clustering~\cite{thaler,delgado_jets,lejarza,deLejarza:2022vhe}, elementary particle process integration~\cite{deLejarza:2023qxk, Agliardi:2022ghn,Williams:2025hza}, anomaly detection~\cite{Belis:2023atb,Schuhmacher:2023pro,kyriienko2022unsupervised}, data classification~\cite{Belis:2024guf, Belis:2021zqi} etc. QML was also used to analyze the causal structure of multi-loop Feynman diagrams and integrating them to predict decay rates at higher orders in perturbation theory~\cite{Ramirez-Uribe:2021ubp,Clemente:2022nll,Ramirez-Uribe:2024wua,deLejarza:2024pgk,deLejarza:2024scm,Pyretzidis:2025stx}. Finally, an increasing part of QML studies is targeting generative modeling and preparing quantum circuits that can mimic relevant probability distributions~\cite{Romero2021,Paine2021,Zhu2022,Kyriienko2024,kasture2022protocols,rudolph2023trainability,Wu2024QHT}. 

Fragmentation Functions (FFs) represent an essential component in describing quantum particle processes, encoding the transition from partons --- quarks and gluons ---  to hadrons after a hard-scattering event~\cite{deFlorian:2007ekg,Aidala:2010bn,deFlorian:2007aj}. Unlike partonic cross-sections, FFs cannot be predicted perturbatively and must be obtained from experimental data via global analysis across diverse processes~\cite{Rojo:2015acz,Albino:2008aa}. Several approaches were proposed to extract FFs from experimental data using statistical tools and heuristics, for instance a polynomial-based ansatz in the hadron's momentum fraction~\cite{Gluck:1998xa,Kretzer:2000yf,Kniehl:2000fe,Hirai:2007cx}, and Euler Beta function distributions with additional parameters~\cite{deFlorian:2007aj,deFlorian:2007ekg,Aidala:2010bn}. Despite the progress, determination of FFs can still be affected by procedural bias, including limited functional form flexibility and uncertainty estimation challenges. Machine Learning (ML) techniques, particularly based on neural networks, have emerged to potentially overcome these limitations~\cite{Nocera:2017qgb, Bertone:2017xsf, Bertone:2017tyb, Soleymaninia:2022qjf}.
\begin{figure*}[t!]
    \includegraphics[width=\textwidth]{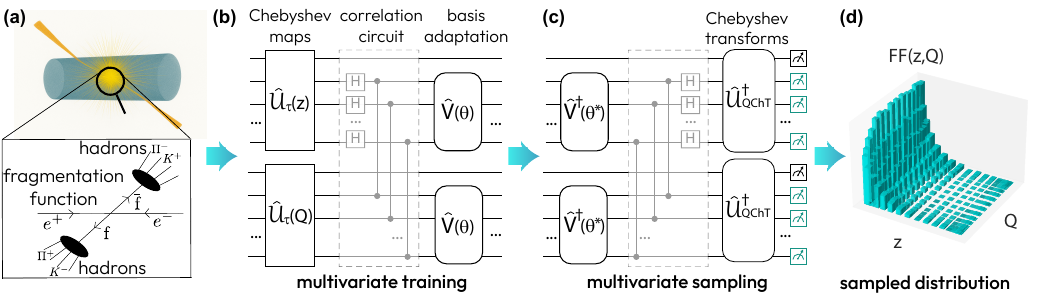} 
    \captionsetup{justification=Justified, singlelinecheck=off} 
    \caption{\textbf{Workflow for describing fragmentation functions with quantum probabilistic models based on shifted Chebyshev polynomials.} (a) The input data for the fragmentation function (FF)    $D_i^\textrm{h}(z,Q)$ is produced for a grid of $z$ and $Q$. (b) Quantum probabilistic model is composed of two Chebyshev feature maps for encoding $z$ and $Q$, a correlation circuit that entangles both registers, and basis adaptation circuits to be trained on $D_i^\textrm{h}(z,Q)$. (c) For sampling we perform the inverse of basis adaptation, the correlation circuit, followed by parallel inverse quantum Chebyshev transforms for mapping the model into the bit basis. (d) Sampling results assembled in a 2-dimensional plot that represents $D_i^\textrm{h}(z,Q)$. }
    \label{fig:workflow}
\end{figure*}
While supervised ML approaches are effective in determining FFs and Parton Density Functions (PDF) in hadronic collisions~\cite{NNPDF:2021uiq}, some challenges remain. 
A primary issue lies in the energy-scale evolution. Models are typically trained for a specific range of momentum fraction $z$ and a fixed energy scale $Q$. However, evolving a given FF to different energy scales requires solving the DGLAP evolution equations~\cite{Gribov:1972ri,Lipatov:1974qm,Altarelli:1977zs,Dokshitzer:1977sg}, a task that is computationally expensive. Additionally, once the FFs are determined, their true value are only known for specific points, necessitating interpolation to estimate the function at other regions, increasing the complexity of the method. Finally, given the inherent nature of FFs as probability distributions, which require inversion to get samples, and the need for multivariate descriptions based on both momentum fraction ($z$) and the energy scale ($Q$), alternative methods are being explored. 

Quantum computing naturally fits generative modeling due to the probabilistic nature of projective measurements and natural sampling abilities from strongly correlated distributions~\cite{nielsen,Arute2019}. Following the Born rule, the probability of measuring a bitstring $x$ is $p(x) = |\langle x| \psi \rangle|^2$, where a wavefunction $| \psi \rangle$ is prepared by some parametrized quantum circuit. We refer to models that represent such circuit-based distributions as quantum probabilistic models of an implicit type~\cite{Kyriienko2024,kasture2022protocols}. These models can be sampled, offering generative modeling capabilities as commonly needed in ML~\cite{Bond2022}. Quantum generative modeling was demonstrated with Quantum Generative Adversarial Networks (QGANs)~\cite{Lloyd2018,Dallaire-Demers2018,Zoufal2019,Huang2021PRAppl}, Quantum Boltzmann Machines (QBMs)~\cite{Amin2018,Zoufal2021,Coopmans2024}, and Quantum Circuit Born Machines (QCBMs)~\cite{Liu2018,Benedetti2019npj,Benedetti_2020,Coyle2020,Kiss2022,hibat2024framework}. To date, quantum generative modeling has shown intriguing results for HEP analysis~\cite{Delgado:2022tpc,Delgado:2023ofr,Delgado:2024vne,Bermot:2023kvh, Tuysuz:2024hyl, Rudolph:2023iqf}. Also, quantum-inspired approaches based on tensor networks are explored as probabilistic models to analyze multivariate distributions ~\cite{carrazza,Cruz-Martinez:2023vgs,Kang:2025xpz}. 

Both Fourier and Chebyshev expansions are common methods for approximating smooth functions using global basis functions. Fourier expansions, which use the complex exponential form of sinusoidal functions, are ideal for periodic functions on uniform grids. In contrast, Chebyshev expansions, which utilize Chebyshev polynomials, are particularly suited for non-periodic functions defined on finite intervals. In our recent works~\cite{williams2023quantum, Wu2024QHT}, we have developed quantum Chebyshev and Hartley  feature map circuits. These circuits are designed to encode input data as real-valued quantum states associated with real expansion coefficients. We have also built their corresponding transform circuits responsible for basis mappings, which are essential for generative modeling. For this specific study, we are focusing primarily on Chebyshev expansions. We have chosen this approach because it offers the best polynomial approximation to a continuous function under the maximum norm, making it particularly well-suited for approximating the fragmentation functions (FFs) relevant to our work.

In this work, we propose a multivariate quantum probabilistic model that  represents HEP data involving several correlated variables, and apply it to describe FFs for the hadronization process. Specifically, we develop Quantum Chebyshev Probabilistic Models (QCPMs) with shifted Chebyshev feature maps and a correlation circuit for multivariate probability distributions~\cite{williams2023quantum,Wu2024QHT}. Our approach allows encoding models \textit{explicitly} in the basis of orthogonal Chebyshev polynomials, and transform trained QML models of variables $z$ and $Q$ into the bivariate distribution $p(z,Q)$ ready for sampling. Such approach provides an ability to infer correlations between variables, giving an insight into HEP data, as well as augmenting HEP datasets with generative modeling on extended grids that grow exponentially with the system size. 
While classical approaches such as~\cite{Bertone:2017xsf} focus on fitting frameworks that provide statistically robust estimations of FF uncertainties, our approach differs in scope and goals. We do not aim to quantify uncertainties; instead, we go beyond traditional polynomial parametrizations of FFs. Using the Chebyshev feature map, we build a more flexible and expressive representation that can capture complex structures missed by standard fits. Furthermore, our method allows efficient interpolation, with resolution growing exponentially with the number of qubits. QPCM training uses small circuits to keep optimization tractable. After training, we exploit the exponential growth of Hilbert space with qubit count to sample at higher resolutions. This separation of training and sampling demonstrates QPCM’s interpolation ability without exceeding classical simulation limits, highlighting its potential for efficient, high-resolution sampling with low training cost. The small number of qubits required for both training and sampling makes hardware execution feasible and also allows efficient simulation using tensor networks. However, scaling the sampling to a larger number of qubits for extremely high resolution would require a bond dimension that grows exponentially, which may limit the performance of the tensor network implementation.

\section*{Results and Discussion}
\subsection{I. Fragmentation functions and hadronization} Our goal is to study HEP processes which: 1) lead to correlated data; 2) require working with multivariate distributions that are hard to study classically; 3) benefit from native abilities of quantum devices for sampling non-trivial probability distributions~\cite{Arute2019}. Here, we concentrate on parton hadronization. These processes are described by fragmentation functions, being two-dimensional probability distributions that possess the properties outlined above. In practice, FFs define differential cross-sections for processes that lead to specific hadrons (e.g. pions or kaons) as a result (Fig.~\ref{fig:workflow}a). For the single-inclusive production of a hadron $h$ in electron-positron annihilation, the differential cross-section is defined from
\begin{equation}
    \frac{d\sigma^\textrm{h}}{dz}(z, Q^2) = \frac{4\pi \alpha^2(Q)}{Q^2} F^\textrm{h}(z, Q^2),
    \label{eq:sigma-main}
\end{equation}
where $F^\textrm{h}(z,Q^2)$ is the fragmentation \textit{structure} function, and $\alpha(Q)$ represents the quantum electrodynamics (QED) running coupling. 

The structure function $F^\textrm{h}(z, Q^2)\propto\sum_{i,j} C_{j}(z, \alpha_\textrm{s}(Q)) \otimes D_i^\textrm{h}(z,Q)$ is defined from a convolution of coefficient functions $C_{j}(z, \alpha_\textrm{s}(Q))$ and fragmentation functions $D_i^\textrm{h}(z,Q)$, where indices $i$ and $j$ denote relevant configurations (e.g. singlet or non-singlet), and $\alpha_\textrm{s}(Q)$ is the QCD running coupling. The convolution operation $\otimes$ is defined as $f(z)\otimes g(z) = \int_z^1 (dy/y) f(y) g(z/y)$. The FFs $D_i^\textrm{h}(z,Q)$ follow the DGLAP evolution equations~\cite{Altarelli:1977zs}, where one needs to perform the evolution of the FFs with the energy scale $Q$ to solve Eq.~\eqref{eq:sigma-main}. As an example, the evolution of the non-singlet component according to the DGLAP equation is given by
\begin{equation}
    \frac{\partial}{\partial \ln{Q^2}} D^\textrm{h}_{\mathrm{NS}}(z, Q^2) = P_+ (z, \alpha_\textrm{s}) \otimes D^\textrm{h}_{\mathrm{NS}}(z, Q^2),
    \label{eq:dglap2-main}
\end{equation}
where $P_+ (z, \alpha_\textrm{s})$ is a splitting function having a perturbative expansion in the strong coupling $ \alpha_\textrm{s}(Q)$ (see  Supplementary Note 1). The integro-differential equation in Eq.~\eqref{eq:dglap2-main} is in general difficult to solve. In our generative modeling, we take a data-driven approach to bypass the DGLAP propagation.

\subsection{II. Quantum Chebyshev Probabilistic Models} We proceed to introduce a framework that is suitable for describing FFs and associated probabilistic processes. The resulting models need to provide access to samples of multivariate distributions yet having a form such that they can be analyzed~\cite{Schuld2021,Kyriienko2021diff}. 
To match the desired requirements, we propose to build QCPMs. These models are based on the quantum Chebyshev toolbox~\cite{williams2023quantum} which allows building models in the basis of orthogonal polynomials. Given the typical shape of distributions, the Chebyshev basis~\cite{Kyriienko2021dqc} fits naturally for describing FFs. We extend this to a generalized setting, specifically developing quantum circuits for two variables, correlating them, and demonstrating sampling capabilities on the extended grids.

The core of QCPM is represented by a probability distribution parametrized as $p(x) = |\langle \tau(x)|\psi\rangle|^2$, where $|\psi\rangle \equiv \hat{V}|0\rangle$ is a quantum state prepared by a unitary circuit $\hat{V}$, and $|\tau(x)\rangle \equiv \hat{\mathcal{U}}_{\mathrm{\tau}}(x)|0\rangle$ is a Chebyshev state generated by the corresponding feature map $\hat{\mathcal{U}}_\mathrm{\tau}$. The Chebyshev state is given by \cite{williams2023quantum}
\begin{align}
\label{eq:tau(x)}
    |\tau(x)\rangle = \frac{1}{2^{N/2}}T_0(x)|0\rangle + \frac{1}{2^{(N-1)/2}} \sum_{k=1}^{2^N-1} T_k(x)|k\rangle,
\end{align}
where $T_{k}(x) \equiv \cos(k \arccos(x))$ is a degree $k$ Chebyshev polynomials of the first kind. These are typically defined in the domain of $[-1,1]$. At the Chebyshev grid points $\{|\tau(x_j)\rangle \}_{j=0}^{2^N-1}$ there is a set of orthonormal states, which are mapped by a unitary transform to the computational basis states, $\{|x_j\rangle \}_{j=0}^{2^N-1}$. Importantly, by applying Chebyshev feature maps in parallel we generate explicit models of several variables, and can introduce correlations by a circuit that connects the registers. This leads to QCPM of the form $p(x,y) = \sum_{k,l=0}^{2^N-1}c_{k,l}T_k(x)T_l(y)$, where $c_{k,l}$ are defined by quantum circuits $\hat{V}$. Our goal is to train such models on FF-based datasets and enable sampling. 

The workflow for QCPM-based HEP analysis is shown in Fig.~\ref{fig:workflow}. In Fig.~\ref{fig:workflow}(a) we visualize a hadronization process, described by the fragmentation function $D_f^\textrm{h}(z, Q)$ and the associated data. These data are transformed into the Chebyshev space and fed into the training [Fig.~\ref{fig:workflow}(b)]. Once training is complete, the inferred model is used for sampling a multivariate back in the problem space [Fig.~\ref{fig:workflow}(c)]. This leads to an augmented FF distribution visualized in Fig.~\ref{fig:workflow}(d).

\begin{figure}[th]
\begin{center}
\includegraphics[width=1.0\linewidth]{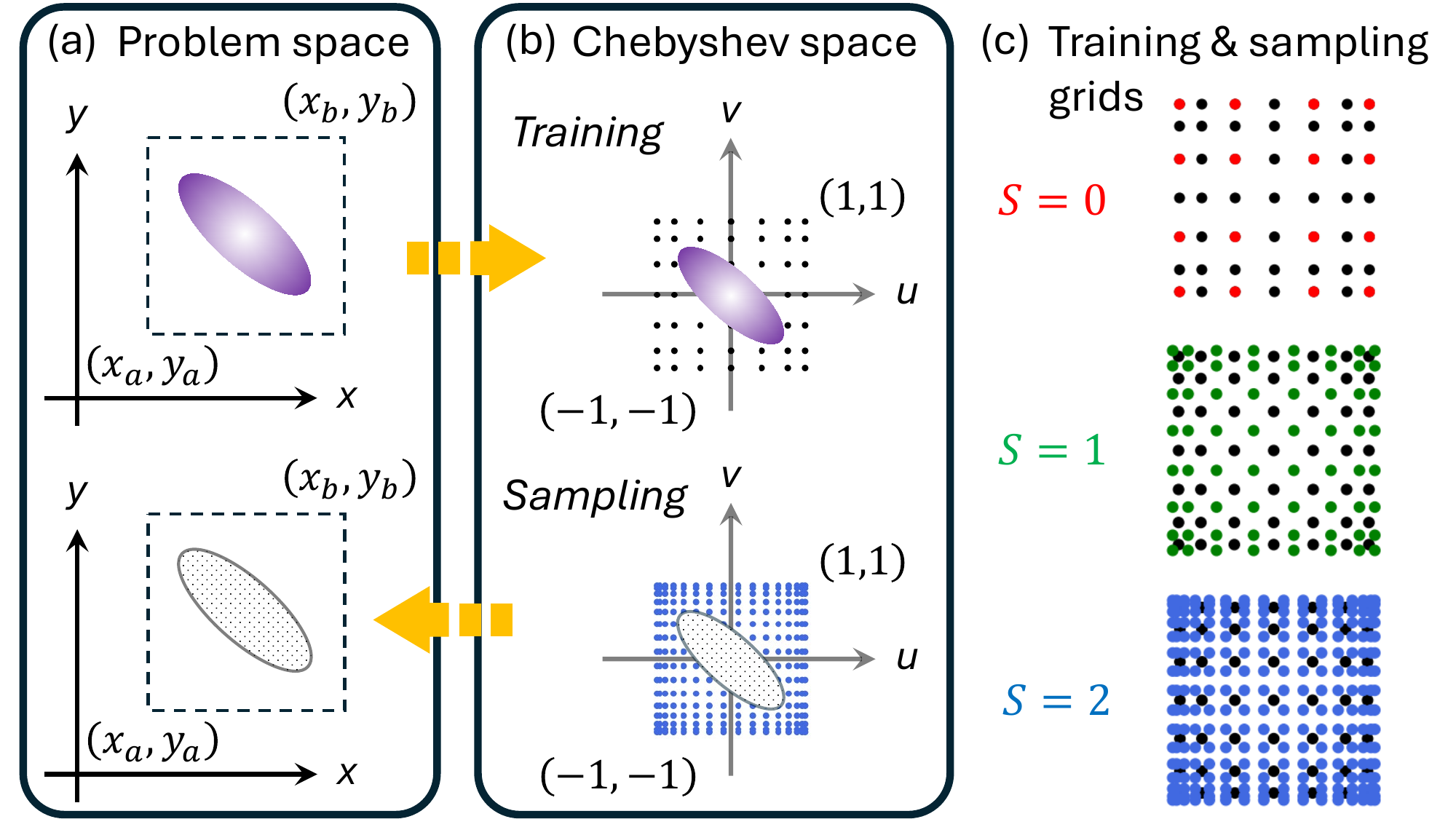}
\end{center}
    \captionsetup{justification=Justified, singlelinecheck=off} 
\caption{\textbf{Applying linear maps and setting up quantum Chebyshev probabilistic models (QCPM).} Application of shifted Chebyshev polynomials to map between (a) the problem and (b) Chebyshev spaces. (c) Visualization of training (black) and sampling (color) grids for different extended registers.}
\label{fig:QCPM-grid}
\end{figure}

As an important methodological detail for FF modeling, we address the challenge of generalizing our models from the original problem domain to the Chebyshev native domain $\Omega_C =[-1,1] \times [-1,1]$ and back. This is illustrated in Fig.~\ref{fig:QCPM-grid}. Given a 2D probability density function $p(x,y)$ defined in a box domain delimited by coordinate tuples, $\Omega_\textrm{P} = [x_\textrm{a},x_\textrm{b}] \times [y_\textrm{a},y_\textrm{b}]$, we express it in the scaled domain as $p(u,v) = \sum_{k,l=0}^{2^N-1}c_{k,l}T_k(u)T_l(v)$, where $u(x)=[2x-(x_\textrm{a}+x_\textrm{b})]/(x_\textrm{b}-x_\textrm{a})$ and $v(y)=[2y-(y_\textrm{a}+y_\textrm{b})]/(y_\textrm{b}-y_\textrm{a})$. The transformation is shown in Fig.~\ref{fig:QCPM-grid}(a and b, top).  
Using the Chebyshev feature maps for the scaled variables $\hat{\mathcal{U}}_\mathrm{\tau}(u)$ and $\hat{\mathcal{U}}_\mathrm{\tau}(v)$, we build the parameterized QCPM $p_{\mathrm{Q}}(u,v)$ 
to approximately represent the transformed probability
density function $p(u,v)$. 
The quantum model is trained on a 2D grid of Chebyshev nodes 
$\{u^{\text{Ch}}_i\}_{i=0}^{2^N-1} \times \{v^{\text{Ch}}_j\}_{j=0}^{2^N-1} $ 
plus additional half-index points $\{u^{\text{Ch}}_{i+1/2}\}_{i=0}^{2^N-2} \times \{v^{\text{Ch}}_{j+1/2}\}_{j=0}^{2^N-2} $. 
The training grid is schematically illustrated in Fig.~\ref{fig:QCPM-grid}(b, top) as a black-dotted grid within $\Omega_C$. 
Once the model is trained, sampling is carried out through projective measurements in the same domain (Fig.~\ref{fig:QCPM-grid}b, bottom), using inverse quantum Chebyshev transform circuits $\hat{\mathcal{U}}_{\mathrm{QChT}}^\dagger$. Details of implementing QCPM are provided in Methods. 

The sampled $p_{\mathrm{Q}}(u,v)$ is mapped back to the problem domain to obtain the sampled histogram $p_{\mathrm{Q}}(x,y)$, rescaled as $x(u)=[(x_\textrm{b}-x_\textrm{a})u+(x_\textrm{a}+x_\textrm{b})]/2$ and $y(v)=[(y_\textrm{b}-y_\textrm{a})v+(y_\textrm{a}+y_\textrm{b})]/2$ and reshaped in a suitable form. This is shown in Fig.~\ref{fig:QCPM-grid}(a, bottom). Crucially, the quantum model is trained on a sparse training grid given by the available training dataset but can provide samples (predictions) for unseen data points on a controllable, denser \textit{sampling grid}, which is not necessarily a superset of the \textit{training grid}. This is based on extending the register to 2$(N+S)$, where $S$ is the number of added qubits on which we act with the Chebyshev transform. In this case, the learned distribution from the training data can closely mimic the physical distribution for the input data, generalizing (interpolating) beyond the training grid. 

In this work, we applied shifted Chebyshev polynomials to study the FFs as a testbed. Shifted Chebyshev polynomials can approximate any continuous function beyond the Chebyshev domain and thus are not limited to FFs. Because the coefficients in the Chebyshev expansion are purely real, it is straightforward that a purely real quantum state should be used. The Chebyshev encoding provides a convenient way to encode each input datum, $v$, as a distinct real quantum state, $\tau(v)$, in terms of orthonormal bases. Most importantly, the proposed QCPMs allow us to systemically and modularly extend to a larger system size and allow us to explore the underlying data correlations inferred by QCPMs. The correlation circuit shown in Fig.~\ref{fig:workflow}(b) is a problem-specific ansatz designed to efficiently train the FFs.  The ansatz circuit can be tailored to fit various applications in different fields.

In addition, the modular capability of QCPM allows us to easily scale up the model dimension at the expense of increasing the circuit complexity mainly contributed by quantum Chebyshev transform circuits. Please refer to our previous work~\cite{williams2023quantum} for details of the quantum circuits.

\subsection{III. Application of QCPM to FFs} We have applied QCPM to FFs by using data accessible via the LHAPDF6 interface~\cite{Buckley:2014ana,lhapdfLHAPDFMain}. The FF datasets are derived from hadron production in electron-positron Single-Inclusive Annihilation (SIA), one of the cleanest processes for studying hadron production, as it does not require simultaneous knowledge of PDFs. 
\begin{figure}[t!]
    \captionsetup{justification=Justified, singlelinecheck=off} 
        \includegraphics[width=1.0\linewidth]{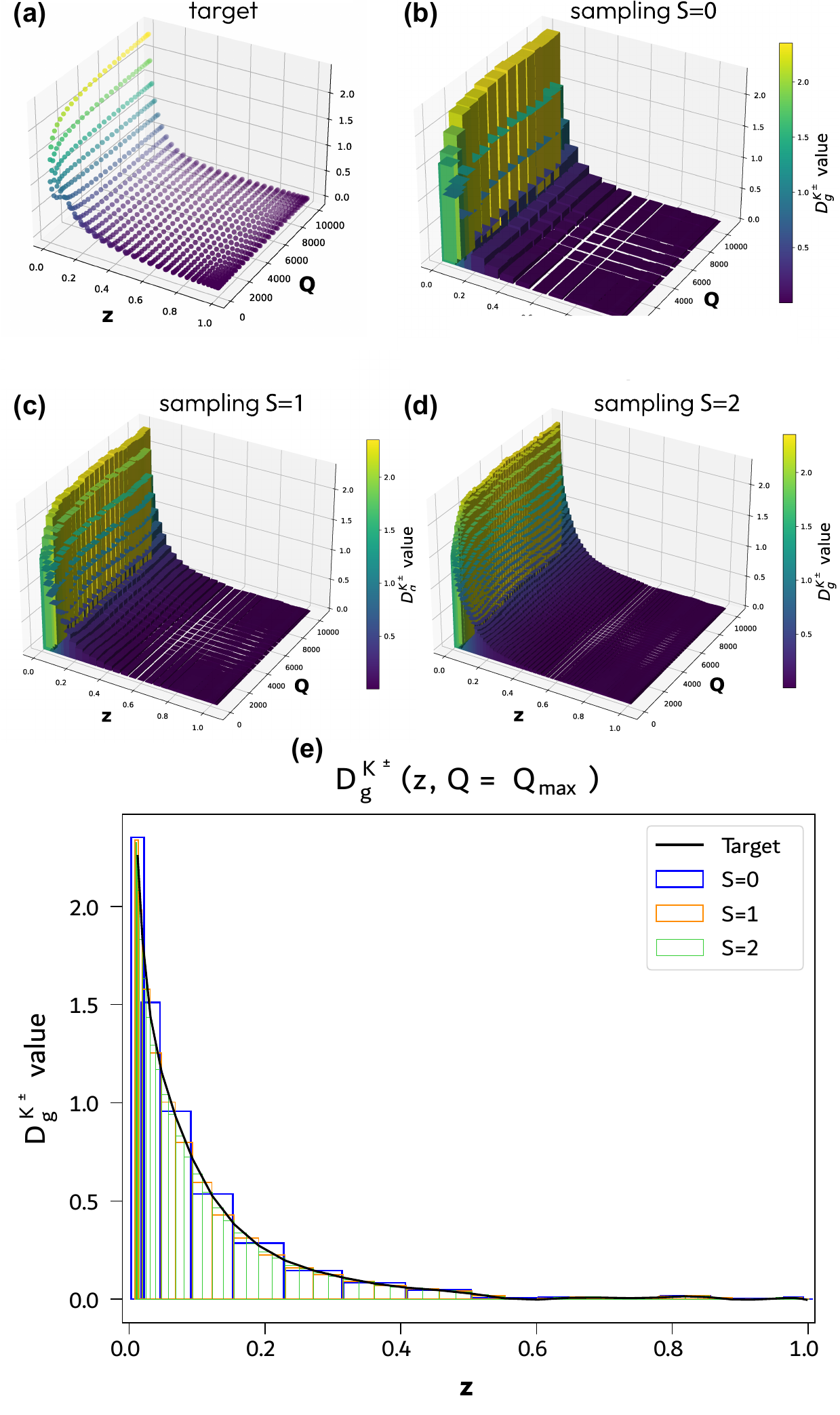}
        \caption{\textbf{Sampling of $D_{g}^{K^\pm}(z,Q)$, the fragmentation function (FF) of a gluon fragmenting into kaons.} In panels (a)-(d) the sampling is performed with $S=0,1,2$ additional qubits for each variable to interpolate in untrained regions. (a) Target distribution $D_{g}^{K^\pm}(z,Q)$. (b) Samples from trained Quantum Chebyshev Probabilistic Model (QCPM) of $D_{g}^{K^\pm}(z,Q)$ with the same number of qubits ($S=0$). (c,d)  Samples from $D_{g}^{K^\pm}(z,Q)$ with extended register ($S=1,2$). (e) Contour overlay of the FF projected onto variable $Q$ with $S = 0$, $1$, and $2$ and the target function. The histogram bins represent an average of the $D_{g}^{K^\pm}$ value for a range of values of $z$. }
    \label{fig:FFresults}
\end{figure}
We use the next-to-next-to-leading order (NNLO) datasets \textsf{NNFF10\_PIsum\_nnlo} for pions ($\pi^\pm = \pi^+ + \pi^-$) and \textsf{NNFF10\_KAsum\_nnlo} for kaons ($K^\pm = K^+ + K^-$)~\cite{Bertone:2017tyb}. 
Regarding the specific FFs analyzed, we consider five independent combinations of FFs for each hadron $h$, $\{D_g^\textrm{h}, D_{b^+}^\textrm{h}, D_{c^+}^\textrm{h}, D_{d^++s^+}^\textrm{h},  D_{u^+}^\textrm{h}\}$, where combined FFs are defined as $D^\textrm{h}_{q+} \equiv D^\textrm{h}_q + D^\textrm{h}_{\bar{q}}$ and $ D_{d^++s^+}^\textrm{h}\equiv D^\textrm{h}_{d^+} + D^\textrm{h}_{s^+} $ for different quarks. The problem domain is set as $\Omega_\textrm{P} = [10^{-2},1] \times [1, 10000]$ GeV for $(z,Q)$.

We present the results of one particular FF as an example of the proposed quantum protocol, $D_{g}^{K^\pm}(z,Q)$, which corresponds to the sum of FFs of the gluon $g$ fragmenting into the kaon $K^+$ and its antiparticle $K^-$. Samplings of other FFs for $h=K^\pm, \pi^\pm$ are available in Fig. S1 and S2.
\begin{figure}[bh!]
    \captionsetup{justification=Justified, singlelinecheck=off} 
        \includegraphics[width=1.0\linewidth]{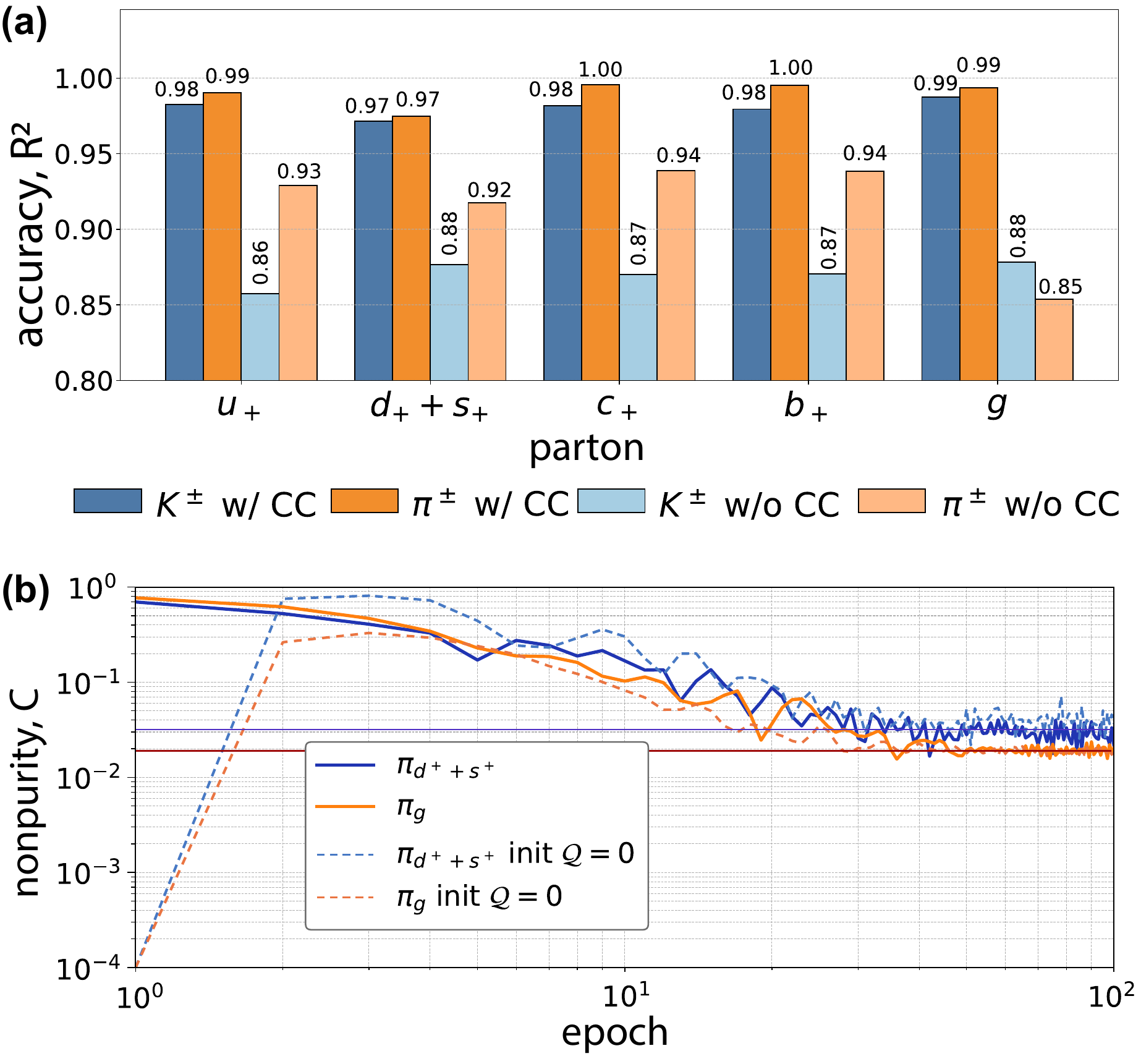}
    \caption{\textbf{Training performance and nonpurity evaluation of the models for different fragmentation functions (FF).} (a) Accuracy ($R^2$) comparison of Quantum Chebyshev Probabilistic Models (QCPMs) for learning FFs $D_i^\textrm{h}(z, Q)$ with (w/ CC) and without (w/o CC) correlations between $z$ and $Q$. (b) Nonpuriry coefficient $C = |1-\gamma_{\mathcal{Z}}|$ of system $\mathcal{Z}$ for 100 training epochs with the fixed correlation circuit (solid curves, log-log curves). Dashed curves represent $C$ when the system starts in a product state without entanglement between registers (no correlations). Solid curves highlight the values of nonpurity after training.}
    \label{fig:results_correlations}
\end{figure}

The results in Fig.~\ref{fig:FFresults} show that QCPM correctly captures the behavior of the 2-dimensional function $D_{g}^{K^\pm}(z,Q)$ in the region of interest. Fig.~\ref{fig:FFresults}(a) depicts the target distribution, shown on the training grid. Fig.~\ref{fig:FFresults}(b) demonstrates the ability of the model to generate samples using the same quantum registers employed during training. Fig.~\ref{fig:FFresults}(c) and Fig.~\ref{fig:FFresults}(d) display the sampling performance when the model utilizes one ($S=1$) and two ($S=2$) additional qubits per variable, respectively, where the model makes predictions in untrained regions, and overall shows excellent correspondence to the ground truth. 

At this point, it is of particular interest to analyze how correlations between variables affect the training process of QCPMs. The introduction of correlations between variables $z$ and $Q$ in the training circuit is motivated by their analytical relationship through the DGLAP evolution equations. 
In this work, we use a heuristic-based approach, where we explicitly introduce a correlation circuit $\hat{C}$, as illustrated in Fig.~\ref{fig:workflow}(b) and detailed in Fig.~\ref{fig:2DQCs}, to entangle the registers that load $z$ and $Q$, which we refer to as $\mathcal{Z}$ and $\mathcal{Q}$. This circuit combines Hadamard gates ($H$) applied to the first variable with controlled-$Z$ gates ($CZ$). Our aim is to evaluate the impact of these correlations on QCPM performance and infer correlations. 

We compare the accuracy of the models with (w/ CC) and without (w/o CC) correlations between the variables, all conditions being equal (following Fig.~\ref{fig:FFresults} and with the same number of qubits, trainable parameters and optimizer epochs). We use the coefficient of determination $R^2$ as a metric to evaluate the goodness of the fit for the explicit QCPMs after the performed training. The results in Fig.~\ref{fig:results_correlations}(a) demonstrate that QCPM with the correlation circuit $\hat{C}$ consistently outperform those without, across all the FFs studied. 
These findings point to the valuable insight from the developed quantum probabilistic models --- we can infer (indirectly) the degree of correlations between physical variables by studying the performance of the quantum model and the need for entanglement between $\mathcal{Z}$ and $\mathcal{Q}$. Although typically correlations are inferred from samples and require significant dataset sizes, we suggest quantifying correlations by unraveling the entanglement properties between registers of QCPMs. This can be achieved by various means (entanglement entropy, purity test, mutual information), and we concentrate on purity measurement \cite{Scali2024} as the one that can be readily obtained from the SWAP test \cite{Paine2022QuantumEquations}. 

Specifically, to quantify the cross-variable correlations we define the nonpurity as $C=|1-\gamma_{\mathcal{Z}}|$, where $\gamma_\mathcal{Z}(\rho)\equiv\textrm{Tr}_{\mathcal{Q}}(\rho_\mathcal{Z}^2)$ is the purity of the subsystem $\mathcal{Z}$, with $\mathcal{Q}$ being traced out. We track the value of $C$ during the sampling stage for QCPM that includes the fixed correlation circuit [c.f. Fig.~\ref{fig:workflow}(c)], and show the change of $C$ for hundred epochs in Fig.~\ref{fig:results_correlations}(b) [two solid curves for selected processes]. Note that the correlations observed in all the FFs are very similar, as can be seen in Fig. S1 and~S2. 
Therefore, here we focus on two FFs for illustrative purposes. We observe that from a large value of $C_0 \approx 0.7$ during the learning stage the model settles nonpurity to a small but finite value $C_{100} \approx 2\times10^{-2}$ [red and blue lines in Fig.~\ref{fig:results_correlations}(b) as a guide]. Note that same values of $C$ are achieved with different circuit initialization chosen to start with uncorrelated registers [dashed curves in Fig.~\ref{fig:results_correlations}(b)]. While being modest, this value reflects the presence of non-negligible $z$-$Q$ correlations that are required to reach high accuracy, as shown in Fig.~\ref{fig:results_correlations}(a). As such, the analysis of QCPM trained on multidimensional datasets leads to quantifiable measures of cross-correlations to high accuracy.

\section*{Conclusions} We introduced quantum Chebyshev probabilistic models (QCPMs) for learning on multivariate HEP distributions, inferring relevant properties, and performing generative modeling by sampling of quantum circuits. We applied QCPMs to describe fragmentation functions (FFs) being probability distributions that describe hadronization processes during high-energy collisions. The developed models in the Chebyshev basis over a generalized grid are particularly suitable for describing FFs that are known to exhibit a polynomial dependence on the momentum fraction. Additionally, QCPM models can be pretrained to describe the required marginal distributions, and tuned further to capture correlations. Our results on inference show that introducing entanglement between $(z,Q)$ quantum registers significantly enhances training performance, reinforcing the growing body of evidence that quantum correlations can improve model accuracy. Our results on generative modeling demonstrate excellent generalization of models between training points (interpolation), where Chebyshev models enable fine grid sampling, where quantum registers extended by $S$ qubits lead $2^{2(S-1)}$-fold increase in sampling grid size. This work paves the way for further exploration of quantum computing techniques to gain deeper insights into fundamental distributions in particle physics. \\

\section*{Methods}

\subsection{Implementation details of Quantum Chebyshev Probabilistic Models}

\begin{figure*}[t!]
\begin{center}
\includegraphics[width=0.8\linewidth]{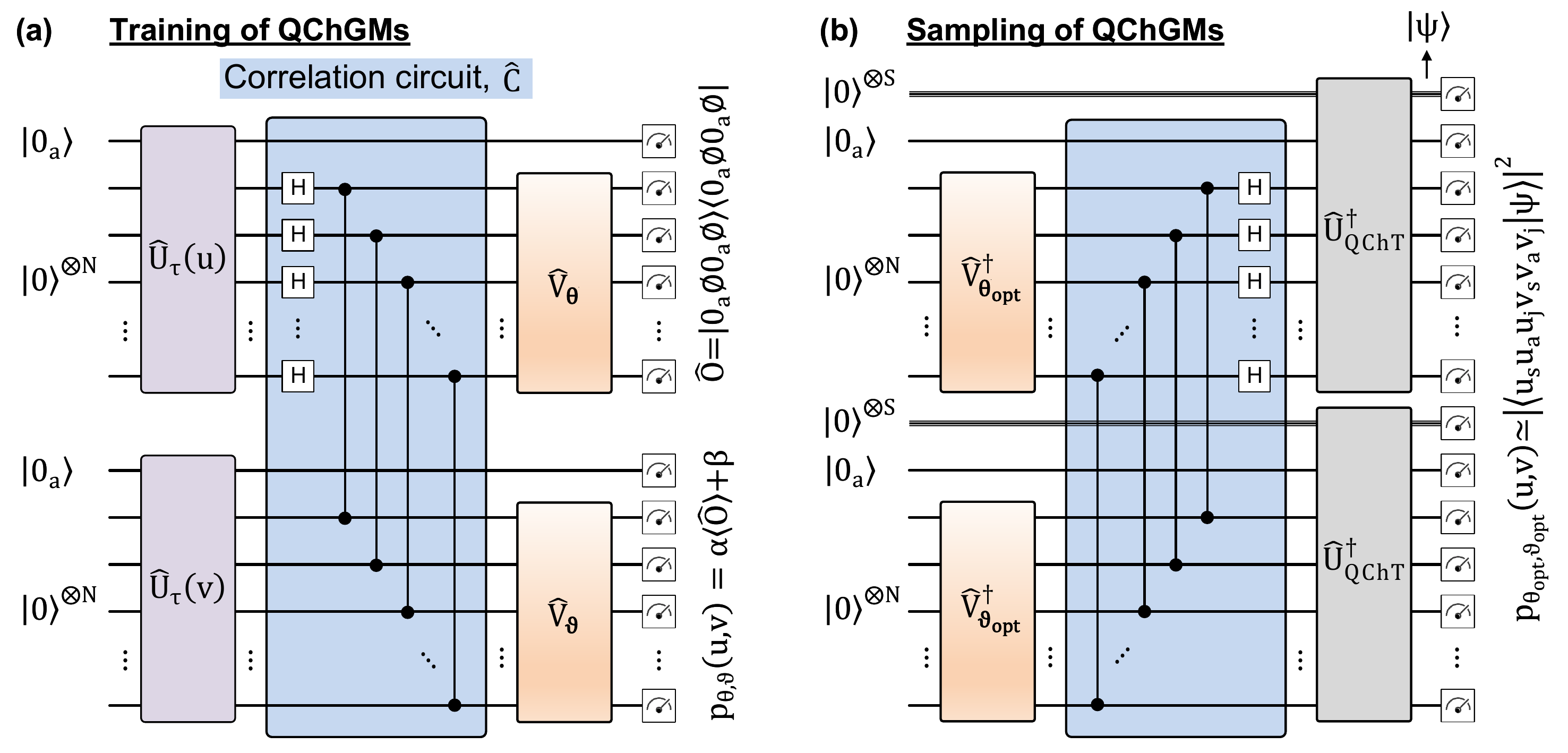}
\end{center}
    \captionsetup{justification=Justified, singlelinecheck=off} 
    \caption{\textbf{Quantum circuits for implementation of QCPMs.}
    (a) Quantum circuit used to train the multivariate distribution in the QCPM latent space, where a correlation circuit $\hat{\mathcal{C}}$ is sandwiched between two identical sets of feature map circuits and variational Ans\"atze. Measured observable is defined as $\hat{\mathcal{O}} = |0_a \mathrm{\o} 0_a \mathrm{\o} \rangle \langle 0_a \mathrm{\o} 0_a \mathrm{\o}|$, where $|\mathrm{\o}\rangle \equiv |0\rangle^{\otimes N}$. Here, $\alpha$ and $\beta$ are trainable scaling and bias parameters. (b) Quantum circuit used to sample the multivariate distribution from the trained model, where the inverse versions of the same parameterized circuits are applied with $\theta^*$ and $\vartheta^*$ being retrieved after the optimization procedure, followed by the inverse versions of the same correlation and two identical sets of quantum Chebyshev transform  circuits associated with extended registers of $S$ qubits ($\ket{0}^{\otimes S}$) in parallel, 
    for fine sampling in the computational basis $|u_s u_\textrm{a} u_j v_s v_\textrm{a} v_j\rangle$. The quantum state prior to measurement is denoted as $|\psi\rangle$.}
\label{fig:2DQCs}
\end{figure*}

\begin{figure}[t]
\begin{center}
\includegraphics[width=1.0\linewidth]{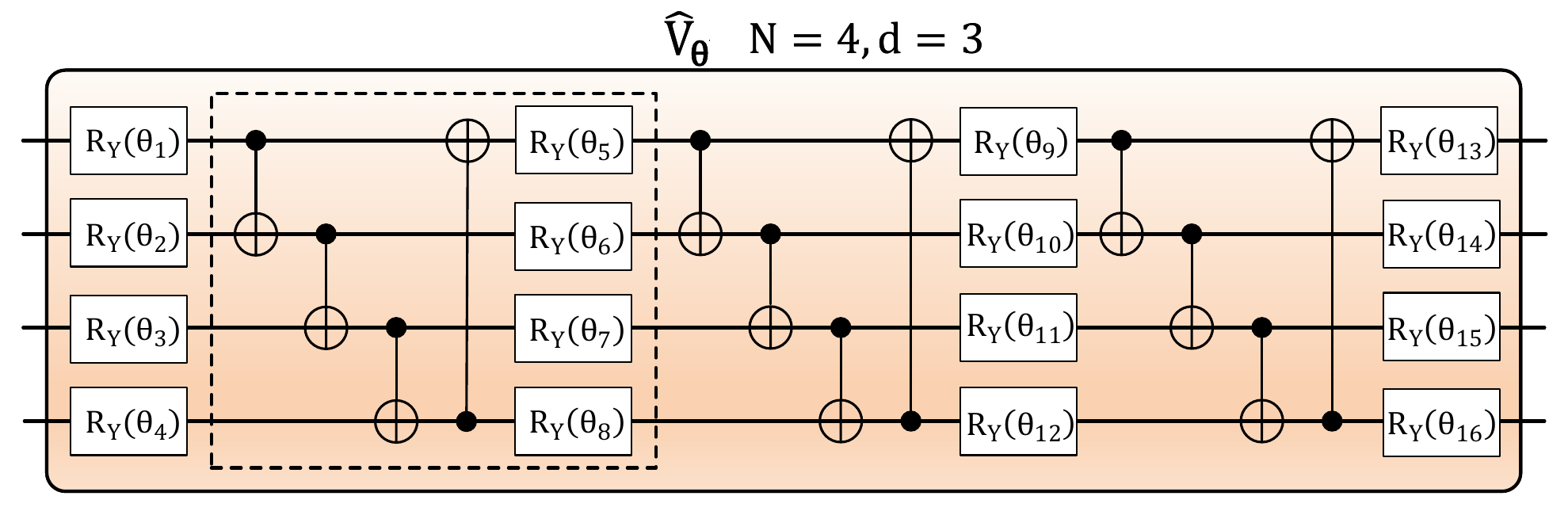}
\end{center}
    \captionsetup{justification=Justified, singlelinecheck=off} 
    \caption{\textbf{Hardware-efficient real-amplitude (HERA) ansatz circuit $\hat{V}_{\theta}$.} The HERA for \textit{N} = 4 qubits with a circuit quantum depth of \textit{d} = 3 is shown. The first block $(d=1)$ is framed by a dashed box. The HERA is composed of $N(d+1)$ tunable single-qubit $\mathrm{R}_\mathrm{Y}$ gates and $N d$ entangling (CNOT) gates. The training parameters are $\{\theta_i\}_{i=1}^{N(d+1)}$.
    }
\label{fig:HERA}
\end{figure}

In this section, we provide implementation details of QCPMs to ensure the reproducibility of our results. The training and sampling quantum circuits employed are presented in Fig.~\ref{fig:2DQCs}.
The training circuit (Fig.~\ref{fig:2DQCs}(a)) is composed of two Chebyshev feature map circuits $\hat{\mathcal{U}}_\mathrm{\tau}(u)$ and $\hat{\mathcal{U}}_\mathrm{\tau}(v)$ for encoding two independent variables, $u$ and $v$ $\in \Omega_C$, in parallel registers, followed by a correlation circuit $\hat{\mathcal{C}}$ and two separate variational Anz\"atze $\hat{\mathcal{V}}(\theta)$ and $\hat{\mathcal{V}}(\vartheta)$. The correlation circuit generates Bell-like entangled states to make two otherwise independent latent variables correlated for efficient training of fragmentation functions. The quantum model $p_{\mathrm{Q}}(u,v) = p_{\theta,\vartheta}(u,v) = \alpha |\langle 0_a \mathrm{\o} 0_a \mathrm{\o} | \bigl( \hat{\text{I}} \otimes \hat{\mathcal{V}}(\theta) \otimes \hat{\text{I}} \otimes \hat{\mathcal{V}}(\vartheta) \bigl) \hat{\mathcal{C}} \bigl( \hat{\mathcal{U}}_\mathrm{\tau}(u) \otimes \hat{\mathcal{U}}_\mathrm{\tau}(v) \bigl) |0_a \mathrm{\o} 0_a \mathrm{\o} \rangle|^2 + \beta$ is trained to search for optimal angles ($\theta^*$ and $\vartheta^*$) and classical weighting parameters ($\alpha_\text{opt}$ and $\beta_\text{opt}$), so that a mean squared error loss function is minimized. Because of  $p_{\theta^*,\vartheta^*}(u,v) \simeq  |\langle u_s u_\textrm{a} u_j v_s v_\textrm{a} v_j| \bigl(\hat{\mathcal{U}}_{\mathrm{QChT}}^\dagger \otimes \hat{\mathcal{U}}_{\mathrm{QChT}}^\dagger \bigl) \hat{\mathcal{C}}_{\text{ext}}^\dagger  \bigl(\hat{\text{I}}^{\otimes (S+a)} \otimes \hat{\mathcal{V}}^{\dagger}(\theta^*) \otimes \hat{\text{I}}^{\otimes (S+a)} \otimes \hat{\mathcal{V}}^{\dagger}(\vartheta^*) \bigl) |0_s 0_a \mathrm{\o} 0_s 0_a \mathrm{\o}\rangle|^2$, the sampling circuit (Fig.~\ref{fig:2DQCs}(b)) consists of the inverse operations of the trained variational Anz\"atze, followed by an extended inverse correlation circuit $\hat{\mathcal{C}}^{\dagger}_{\text{ext}}$ and two identical sets of extended inverse quantum Chebyshev transform $\hat{\mathcal{U}}_{\mathrm{QChT}}^\dagger$ circuits. This procedure allows the basis transformation from the Chebyshev to computational basis spaces. In this context, we assume that the problem domain of interest is expressed in terms of the
momentum fraction and energy scale $(z, Q)$ $\in \Omega_\textrm{P}$.

It is worth mentioning that \Eq{eq:tau(x)} represents the (unnormalized) Chebyshev state. Its normalized version can be prepared by quantum Chebyshev feature map circuit $\hat{\mathcal{U}}_\mathrm{\tau}(u)$ with respect to any continuous input variable $u$. The Quantum Chebyshev transform $\hat{\mathcal{U}}_{\mathrm{QChT}}$ is a basis transformation between the bitstring and the Chebyshev state at the Chebyshev nodes. Details of both quantum circuits are given in our previous work~\cite{williams2023quantum}.

The setup we consider for training the QCPM is as follows. We fix the epoch count of the ADAM~\cite{adam} optimizer to $10^4$, the number of Ansatz layers to 3, and the number of qubits per variable to 4. Hence, both circuits have the same number of free parameters (16 for each variable) and require the same training time. We perform a sweep over the learning rate of the ADAM optimizer across the range $[0.1, 1.0]$ and select the model with the best accuracy in learning the functions $D_i^\textrm{h}(z,Q)$. All quantum simulations are performed using Pennylane \cite{Bergholm:2018cyq}, and the training process is accelerated with JAX \cite{jax2018github}. The Ansatz used for the variational quantum circuit is depicted in Fig.~\ref{fig:HERA}.

We use the Hardware-Efficient Ansatz (HEA) primarily for its simplicity and compatibility with NISQ devices, which facilitates future implementation on quantum hardware. This choice is not motivated by any particular mathematical structure or claim of optimality for the QPCM. More structured circuit families may offer better performance or reduced resource requirements, especially when considering compilation to fault-tolerant gate sets. However, such optimization is outside the scope of this work.

To provide a baseline reference, we include resource estimates for compiling our HEA circuit into a fault-tolerant Clifford+T gate set using the Ross–Selinger decomposition method implemented in \texttt{Pennylane}~\cite{ross-selinger} in Table \ref{tab:hea-rs-resources}. We define the approximation accuracy by the maximum allowable operator norm error $\epsilon$ for the full circuit. The resource estimates include total circuit depth, T-count, and total number of gates. T-count is especially relevant in fault-tolerant quantum computing due to its impact on qubit overhead and magic state distillation costs~\cite{gosset2013algorithmtcount,Postler_2022,Gheorghiu:2021yyp,Bravyi_2005}. For context, recent fault-tolerant proposals report that decomposing a single gate may require on the order of 100 T gates~\cite{Loaiza:2025hfk}, indicating that the T counts we obtain are reasonable for the scale of circuits used in this work.

\begin{table}[h!]
\centering
\caption{Resource estimation for transpiled HEA using Ross–Selinger method}
\label{tab:hea-rs-resources}
\begin{tabular}{|c|c|c|c|}
\hline
\textbf{$\epsilon$} & \textbf{Total Depth} & \textbf{T-count} & \textbf{Total Gate Count} \\
\hline
$10^{-2}$ & 415 & 360 & 1506 \\
$10^{-3}$ & 566 & 512 & 2092 \\
$10^{-4}$ & 716 & 654 & 2681 \\
$10^{-5}$ & 841 & 698 & 2704 \\
\hline
\end{tabular}
\end{table}

Unlike the structure of GAN represented by the simultaneous training of two neural networks, QCPM separates the training and sampling stages. Because the probability distribution is encoded in the orthogonal Chebyshev basis states provided by the quantum Chebyshev feature map circuit, the model effectively represents a wide range of functions with sufficient training data and learns patterns with a decent number of variational parameters. The number of variational parameters used for training in this work is 32 $(= 2 N (d+1))$ with $N$ = 4 and $d$ = 3. The trained model generates new data that fall within the trained distribution.
In addition to generation and sampling, the QCPM efficiently encodes probability distributions into quantum states, making it a valuable tool for a wider range of quantum algorithms and applications.

\section*{Data availability}
The raw data that support the findings of this study are available in \url{https://github.com/CERN-IT-INNOVATION/QChPM}.

\section*{Code availability}
The code and the dataset used in this work is open source and available at \url{https://github.com/CERN-IT-INNOVATION/QChPM}.

\section*{Author Contributions}
JML, GR and MG conceived the initial idea of analyzing fragmentation functions with quantum Chebyshev models. JML generated the data and executed the simulations. O.K. proposed the probabilistic modeling regime for inferring correlations and the multivariate architecture. H.-Y.W. conceived the idea of applying shifted Chebyshev polynomials to the quantum Chebyshev models and provided the initial demo codes. All authors discussed the results and contributed to the final manuscript.

\section*{Competing interests}
Michele Grossi is a Guest Editor for Communications Physics, but was not involved in the editorial review of, or the decision to publish this article. All other authors declare no competing interests

\begin{acknowledgments}
\section*{Acknowledgements} The authors thank Antonio A. Gentile for useful discussions and suggestions on the manuscript. 
This work is supported by the Spanish Government (Agencia Estatal de Investigaci\'on MCIU/AEI/10.13039/501100011033) Grants No.~PID2023-146220NB-I00, No. PID2020-114473GB-I00 and No. CEX2023-001292-S, and Generalitat Valenciana Grants No. ASFAE/2022/009 (Planes Complementarios de I+D+i, NextGenerationEU). This work is also supported by the Ministry of Economic Affairs and Digital Transformation of the Spanish Government and NextGenerationEU through the Quantum Spain project and by the CSIC Interdisciplinary Thematic Platform on Quantum Technologies (PTI-QTEP+). JML is supported by Generalitat Valenciana Grant No. ACIF/2021/219. H.-Y.W. and O.K. acknowledge the funding from UK EPSRC award under the Agreements EP/Z53318X/1 (QCi3 Hub) and EP/Y005090/1. 
MG is supported by CERN through CERN Quantum Technology Initiative.
\end{acknowledgments}

\section*{References}

\bibliography{bibliography}

\clearpage 
\onecolumngrid

\makeatletter
\setcounter{figure}{0}
\renewcommand \thesection{S\@arabic\c@section}
\renewcommand\thetable{S\@arabic\c@table}
\renewcommand \thefigure{S\@arabic\c@figure}
\makeatother

\section*{Supplemental Material}

\subsection{Background on fragmentation functions}

FFs provide the probability that a parton (quark or gluon) fragments into a particular hadron.  
In practice, FFs define the differential cross-section for the single-inclusive production of an hadron $h$ in electron-positron annihilation,
\begin{equation}
    \frac{d\sigma^h}{dz}(z, Q^2) = \frac{4\pi \alpha^2(Q)}{Q^2} F^h(z, Q^2),
    \label{eq:sigma}
\end{equation}
where $F^h(z,Q^2)$ is the fragmentation \textit{structure} function, $z$ is a momentum fraction after scattering, $Q$ is an associated energy scale, and $\alpha(Q)$ represents the quantum electrodynamics (QED) running coupling.

The structure function is defined as a convolution between coefficient functions and FFs:
\begin{equation}
\begin{split}
    \hspace{-0.25cm} F^h(z, Q^2)  &= \frac{1}{n_f} \sum_{q} \hat{e}_q^2  
     \bigg[  C_{{\rm S}2, q}(z, \alpha_s(Q)) \otimes D^h_{\rm S}(z, Q^2) \\
    & + C_{{\rm S}2, g}(z, \alpha_s(Q)) \otimes D^h_g(z, Q^2) +  C_{{\rm NS}2, q}(z, \alpha_s(Q)) \otimes D^h_{\rm NS}(z, Q^2) \bigg],
\end{split}
\end{equation}
where $ \hat{e}_q $ are scale-dependent quark electroweak charge factors, defined in~\cite{deFlorian:1997zj} and $\alpha_s(Q)$ is the QCD running coupling. The sum is performed over the $n_f$ active flavours at the energy scale $Q$, and $C_{{\rm S}2, q}$, $C_{{\rm S}2, g}$, $C_{{\rm NS}2, q}$ are the coefficient functions corresponding respectively to the singlet and
nonsinglet combinations of FFs, 
\begin{align}
    D^h_{\rm S}(z, Q^2) &= \sum_q D^h_{q+}(z, Q^2),  \nonumber \\
    \hspace{-0.15cm} D^h_{\rm NS}(z, Q^2) &= \sum_q \frac{\hat{e}_q^2}{\langle e^2 \rangle} \left[ D^h_{q+}(z, Q^2) - D^h_S(z, Q^2) \right],
    \label{eq:ffns}
\end{align}
and to the gluon FF, $D^h_g(z, Q^2)$. Note that in Eq.~\ref{eq:ffns} the notation $D^h_{q+} \equiv D^h_q + D^h_{\bar{q}}$ and $\langle e^2 \rangle \equiv \frac{1}{n_f} \sum_q \hat{e}_q^2$ has been used. The usual convolution integral with respect to $z$ is denoted by $\otimes$ and reads
\begin{equation}
    f(z)\otimes g(z) = \int_z^1 \frac{dy}{y} f(y) g\left(\frac{z}{y}\right).
\end{equation}

Then, to solve Eq.~\ref{eq:sigma} it is needed to perform the evolution of the FFs with the energy scale $ Q $, which follows the DGLAP evolution equations~\cite{Gribov:1972ri, Lipatov:1974qm, Altarelli:1977zs, Dokshitzer:1977sg}. The singlet component $ D^h_{\rm S}(z, Q^2) $ mixes with the gluon FF, and is given by:
\begin{equation}
\begin{split}
    \frac{\partial}{\partial \ln{Q^2}} \begin{pmatrix} D^h_{\rm S} \\ D^h_g \end{pmatrix} (z, Q^2) & = \begin{pmatrix} P_{qq} & 2n_f P_{gq} \\ P_{qg} & P_{gg} \end{pmatrix} (z, \alpha_s)  \\ & \otimes \begin{pmatrix} D^h_{\rm S} \\ D^h_g \end{pmatrix} (z, Q^2),
    \label{eq:dglap1}
\end{split}
\end{equation}
while the non-singlet component $ D^h_{\rm NS}(z, Q^2) $ evolves as:
\begin{equation}
    \frac{\partial}{\partial \ln{Q^2}} D^h_{\rm NS}(z, Q^2) = P_+ (z, \alpha_s) \otimes D^h_{\rm NS}(z, Q^2),
    \label{eq:dglap2}
\end{equation}
with $P_+=P_{qq}+P_{q\bar q}$ and the splitting functions $P_{ji}$ having a perturbative expansion in the strong coupling $ \alpha_s $:
\begin{equation}
    P_{ji}(z, \alpha_s) = \sum_{l=0} a_s^{l+1} P^{(l)}_{ji}(z),
\end{equation}
where $ j, i = g, q $, and $ a_s = \alpha_s/4\pi $. The time-like splitting functions are known up to $\mathcal{O}(a^3)$ in the modified minimal subtraction, or $\overline{\textrm{MS}}$, scheme~\cite{Almasy:2011eq,Moch:2007tx,Mitov:2006ic}.

momentum fraction and energy scale $(z, Q)$ $\in \Omega_P$.


%

\subsection{Training and sampling of fragmentation functions of kaon and pion}

\begin{figure}[h]
\includegraphics[width=.99\textwidth]{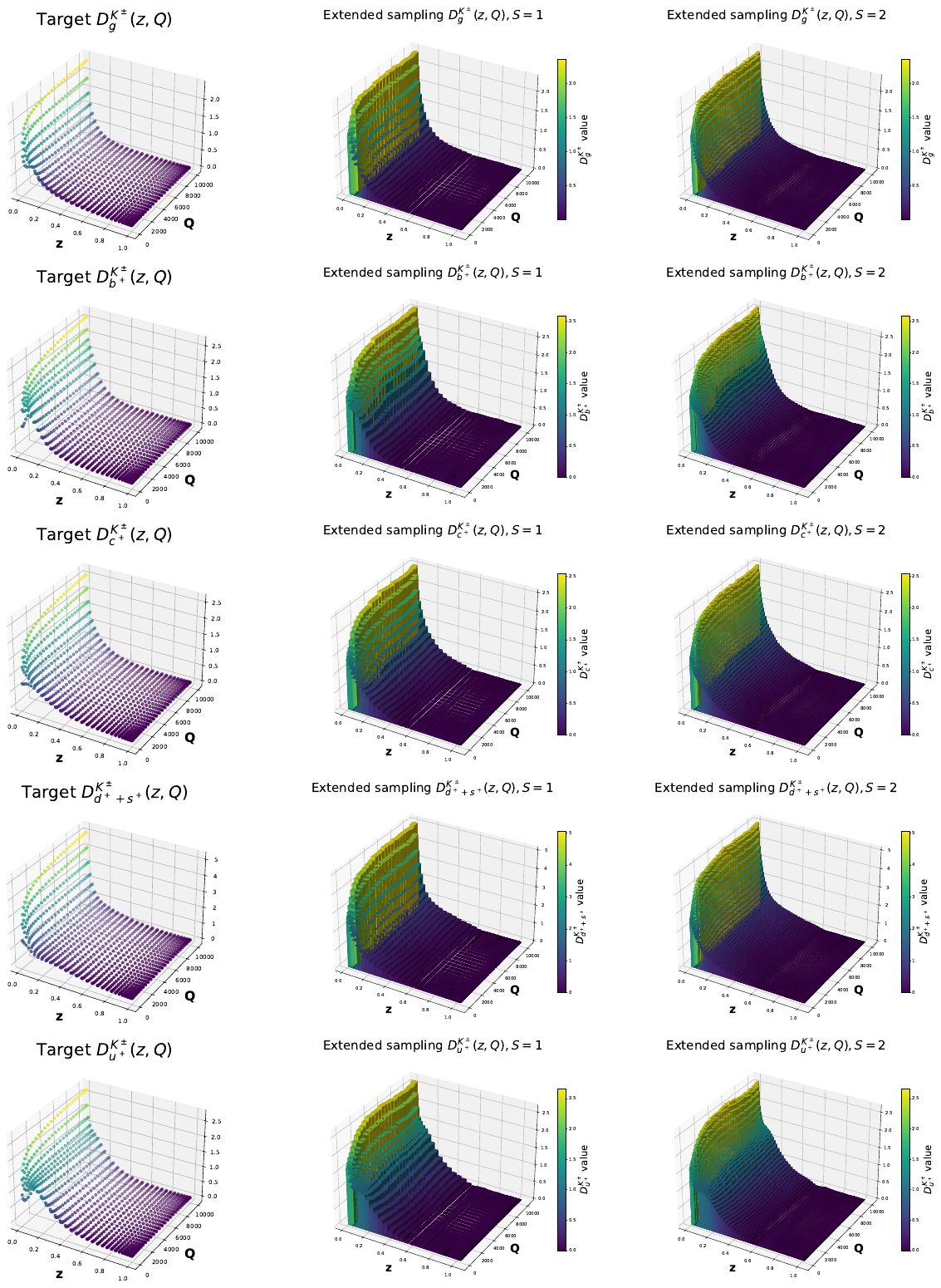}
\caption{Sampling of fragmentation functions of partons into kaons $K^+$ and $K^-$.}
\label{fig:FF_KA_results}
\end{figure}

\begin{figure}[h]
\includegraphics[width=.99\textwidth]{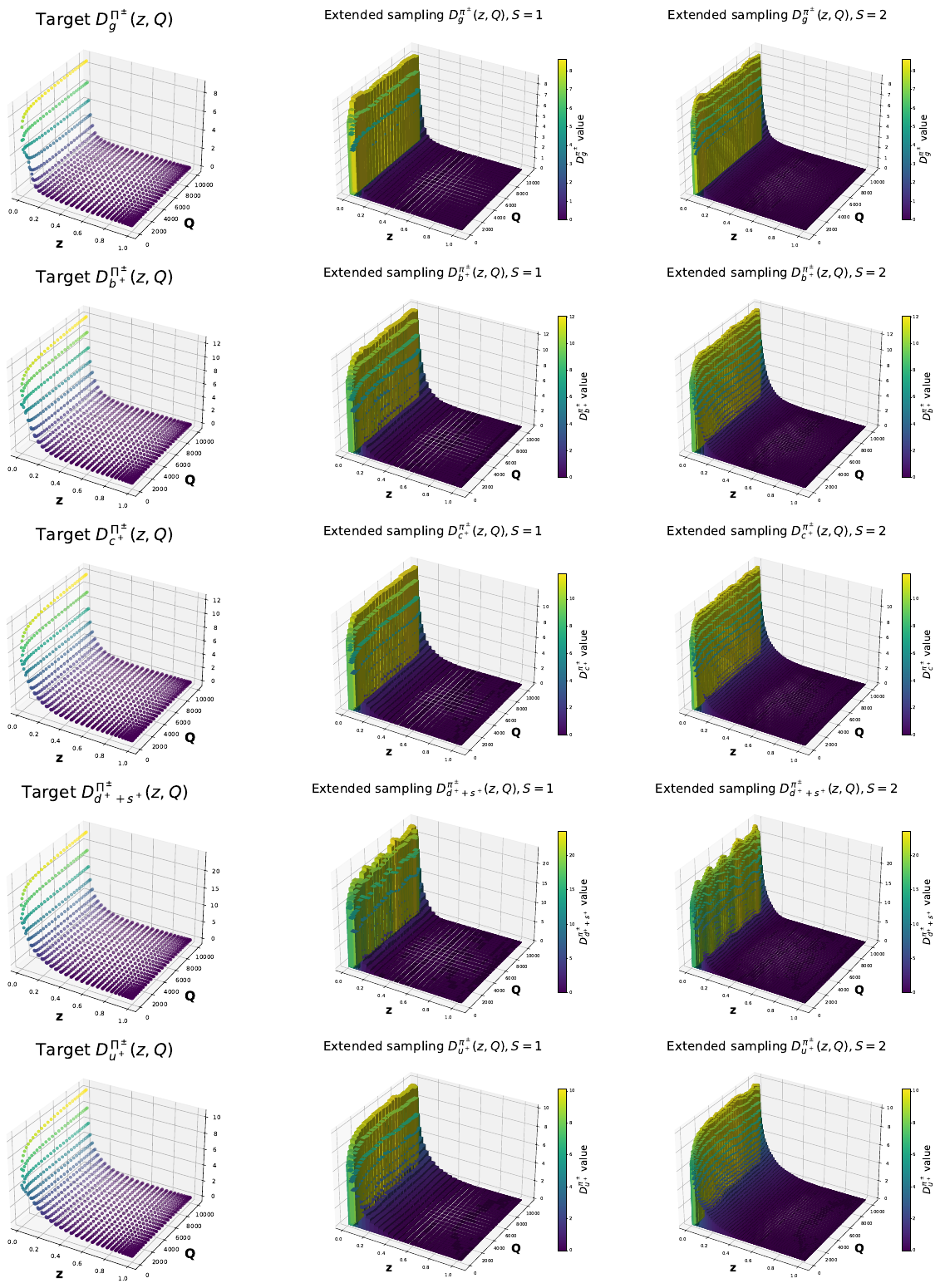}
\caption{Sampling of fragmentation functions of partons into pions $\pi^+$ and $\pi^-$.}
\label{fig:FF_PI_results}
\end{figure}

In this section, we provide a visualization of training (implicitly) and sampling (explicitly) of all FFs considered for partons fragmenting into kaons and pions. In particular, the ten FF distributions considered are:
\begin{equation}
\begin{split}
\{ D_g^{h}, D_{b^+}^{h}, 
D_{c^+}^{h}, 
D_{d^++s^+}^{h} 
D_{u^+}^{h} \}, \qquad  h = K^\pm, \pi^\pm~,
\end{split}
\end{equation}
where $ D^h_{q+} \equiv D^h_q + D^h_{\bar{q}} $, and $ D_{d^++s^+}^h\equiv D^h_{d^+} + D^h_{s^+} $. Also $D_i^{h^\pm}=D_i^{h^+}+D_i^{h^-}$, with $i$ being the fragmenting parton. 
In Fig.~\ref{fig:FF_KA_results} and \ref{fig:FF_PI_results} we show how the quantum model successfully learns all the FFs distributions studied and generates samples out of them.

Regarding the resources used for sampling, we have employed $10^6$ shots to obtain enough data points to provide a smooth description of the target function that correctly mimics the FF behavior. When performing extended sampling with additional qubits for each variable the number of qubits is increased $2S$ times, expanding the Hilbert space. Specifically, for $S=1$, the Hilbert space increases by a factor of~$4$, and for $S=2$, it expands by a factor of~$16$. This increase in the Hilbert space translates to a need for 4 times more shots when using $S=1$ and 16 times more shots when using $S=2$ to maintain the same level of accuracy. Additionally, this enlargement of the Hilbert space results in smaller measurement bins, with the area of each bin becoming 4 times smaller for $S=1$ and 16 times smaller for $S=2$.

\subsection{Von Neumann entropy in quantum registers}

In this section, we quantify the entanglement within both registers encoding $z$ and $Q$, in the training stage of Fig. \ref{fig:2DQCs}(a), using the von Neumann entropy. The von Neumann entropy is a fundamental measure of entanglement for pure bipartite states in quantum information theory \cite{Neumann1927, nielsen}. For a quantum state described by a density matrix $\rho$, the von Neumann entropy is defined as
\begin{equation}
S(\rho) = -\text{Tr}(\rho \log_2 \rho) = -\sum_i \lambda_i \log_2 \lambda_i~,
\end{equation}
where $\lambda_i$ are the eigenvalues of $\rho$. For a pure state $|\psi\rangle$, the von Neumann entropy of either subsystem quantifies the entanglement between the subsystems \cite{Bennett1996}. In the context of bipartite systems, the entropy of entanglement is given by
\begin{equation}
S(\rho_A) = S(\rho_B) = -\text{Tr}(\rho_A \log_2 \rho_A) = -\text{Tr}(\rho_B \log_2 \rho_B)~,
\end{equation}
where $\rho_A$ and $\rho_B$ are the reduced density matrices of subsystems $A$ and $B$, respectively~\cite{Hill1997}. In our analysis, we focus on two primary systems, each encoded by $N=4$ qubits. System $\mathcal{Z}$ that encodes the variable $z$ and system $\mathcal{Q}$ that encodes the variable $Q$. To investigate internal entanglement, we divide each system into two equal subsystems of $N/2=2$ quits each. This partitioning allows us to analyze the entanglement within each system independently.

\begin{figure}[t]
\includegraphics[width=.8\textwidth]{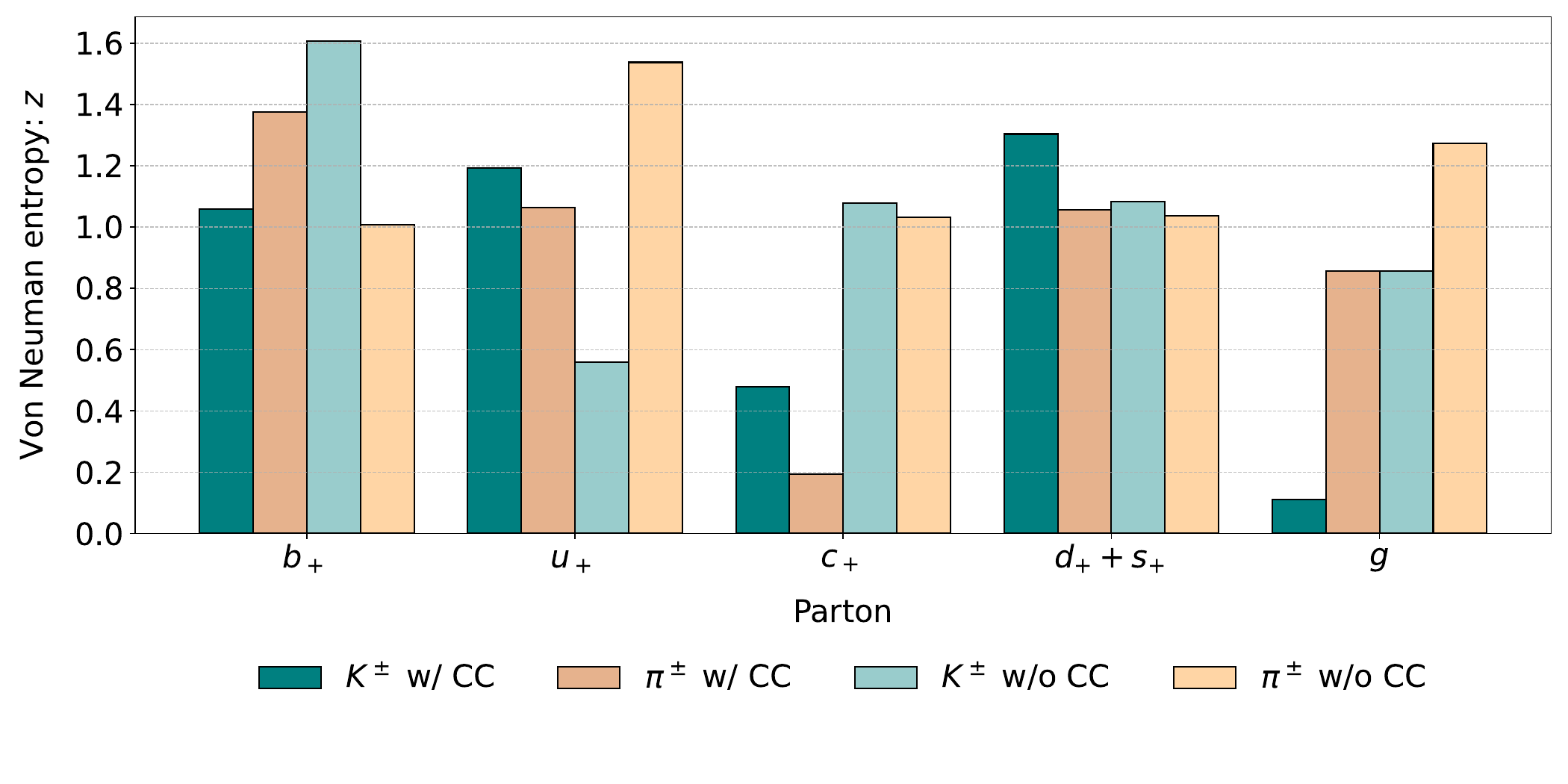}
 \includegraphics[width=.8\textwidth]{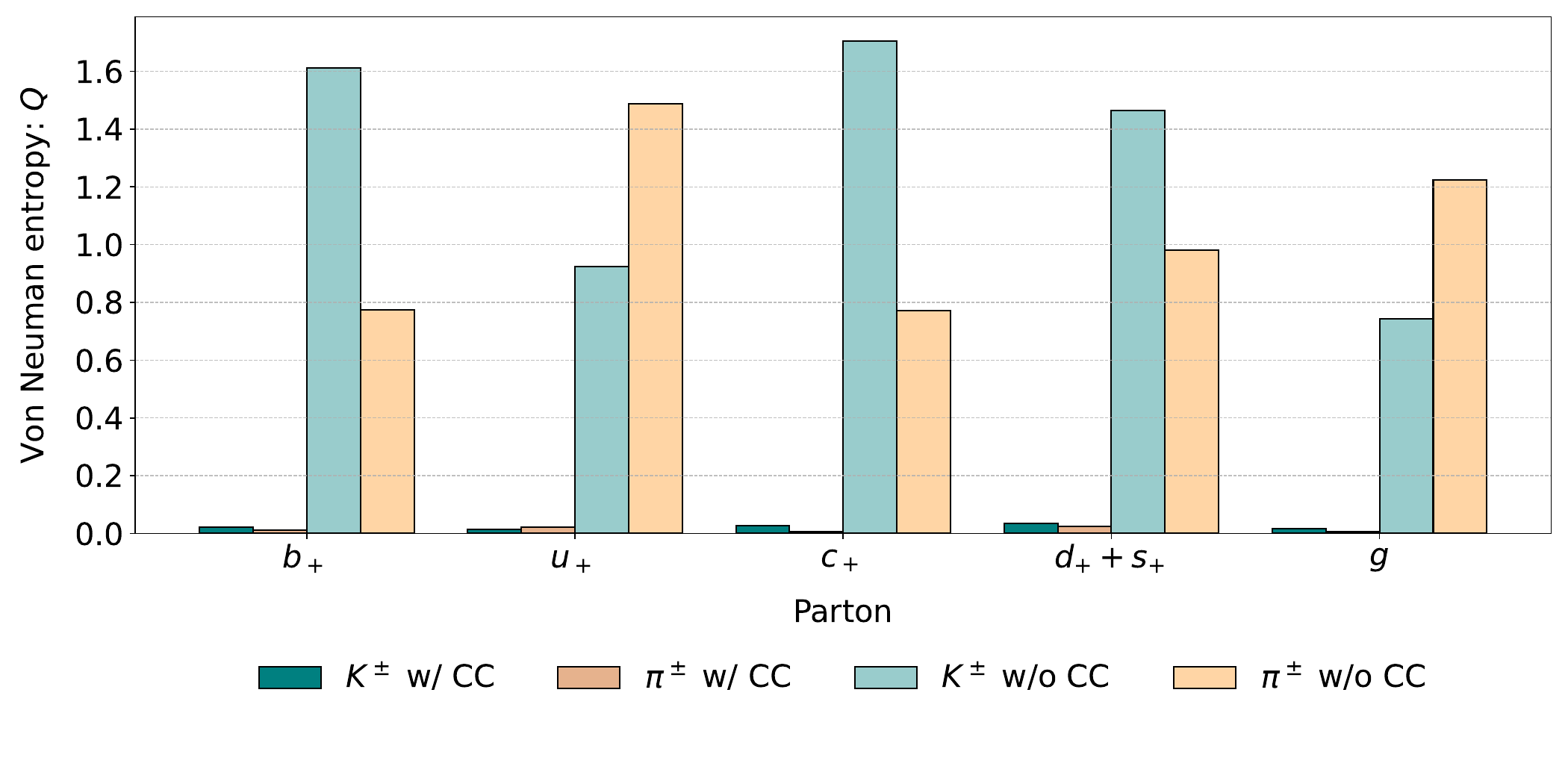}
\caption{Von Neumann entropies of the circuits in $z$ registers (top) and $Q$ registers (bottom) with (w/ CC) and without (w/o CC) correlations between $z$ and $Q$.}
\label{fig:vonneuman}
\end{figure}

It is important to note that we do not calculate the von Neumann entropy between systems $\mathcal{Z}$ and $\mathcal{Q}$ directly. These systems are entangled using $CZ$ gates, which would result in zero entropy when considered as a whole. Instead, our focus is on the internal entanglement within each system.
For the calculation of von Neumann entropies of systems $\mathcal{Z}$ and $\mathcal{Q}$, we utilize the \texttt{qml.math.vn\_entropy} function provided by PennyLane \cite{vn_entropy}. The results of our entropy calculations are presented in Fig. \ref{fig:vonneuman}.


The results presented in Fig. \ref{fig:vonneuman}(a) show significant variations in the von Neumann entropy within the $\mathcal{Z}$ system across the different circuits that have learned the FFs. However, these variations do not appear to correlate with the inclusion or exclusion of the correlations circuit between systems $\mathcal{Z}$ and $\mathcal{Q}$. This suggests that while the internal entanglement structure of the $\mathcal{Z}$ system is dynamic, it is not driven by the inter-system correlations.

On the other hand, the results of Fig. \ref{fig:vonneuman}(b) reveal a strong dependence of the von Neumann entropy in the $\mathcal{Q}$ system on the inclusion of the correlations circuit. When the correlations circuit is not present, we observe a high von Neumann entropy, indicating a substantial level of internal entanglement within the $\mathcal{Q}$ system.  Nonetheless, when the correlations circuit is included, the von Neumann entropy drops drastically, approaching almost zero. This significant reduction in entropy suggests a transfer or ``absorption'' of entanglement from the $\mathcal{Q}$ system to the $\mathcal{Z}$ system. Such behavior aligns with the concept of entanglement monogamy, where increased entanglement between two systems can lead to decreased internal entanglement within one of the systems \cite{Zong_2022}.\\

\subsection{Purity in quantum registers}

In quantum information theory, purity is a measure of the mixedness of a quantum state. For a quantum system described by a density matrix $\rho$, the purity is defined as:
\begin{equation}
\gamma(\rho) = \text{Tr}(\rho^2)~,
\end{equation}
where $\rho$ is the density matrix of the system. A purity of $\gamma(\rho) = 1$ corresponds to a pure state, meaning the system is in a definite quantum state. In contrast, if $\gamma(\rho) < 1$, the state is mixed, indicating the system is in a probabilistic combination of different states. In the case of a maximally mixed state, $\gamma(\rho) = 1/d$, where $d$ is the dimension of the Hilbert space \cite{Jaeger2007-JAEQIA}. The purity of a quantum system like a quantum circuit provides valuable information about the degree of coherence or entanglement in the quantum register. The higher the mixing the higher the entanglement and vice-versa. 

In this section, we quantify the entanglement between the registers encoding $z$ and $Q$, in the training stage of Fig. \ref{fig:2DQCs}(a), using their purity. In particular, we calculate the purity of the system $\mathcal{Z}$, of $N=5$ qubits, that encodes the variable $z$ which by construction is equivalent to the system $\mathcal{Q}$ that encodes the variable $Q$. This purity is computed by tracing out the system that we are not interested in, 
\begin{equation}
    \gamma_\mathcal{Z}(\rho)\equiv\textrm{Tr}(\rho_\mathcal{Z}^2)=\textrm{Tr}(\rho_\mathcal{Q}^2)\equiv\gamma_\mathcal{Q}(\rho).
    \label{eq:purities}
\end{equation}

\begin{figure}[h]
    \centering
    \includegraphics[width=0.5\linewidth]{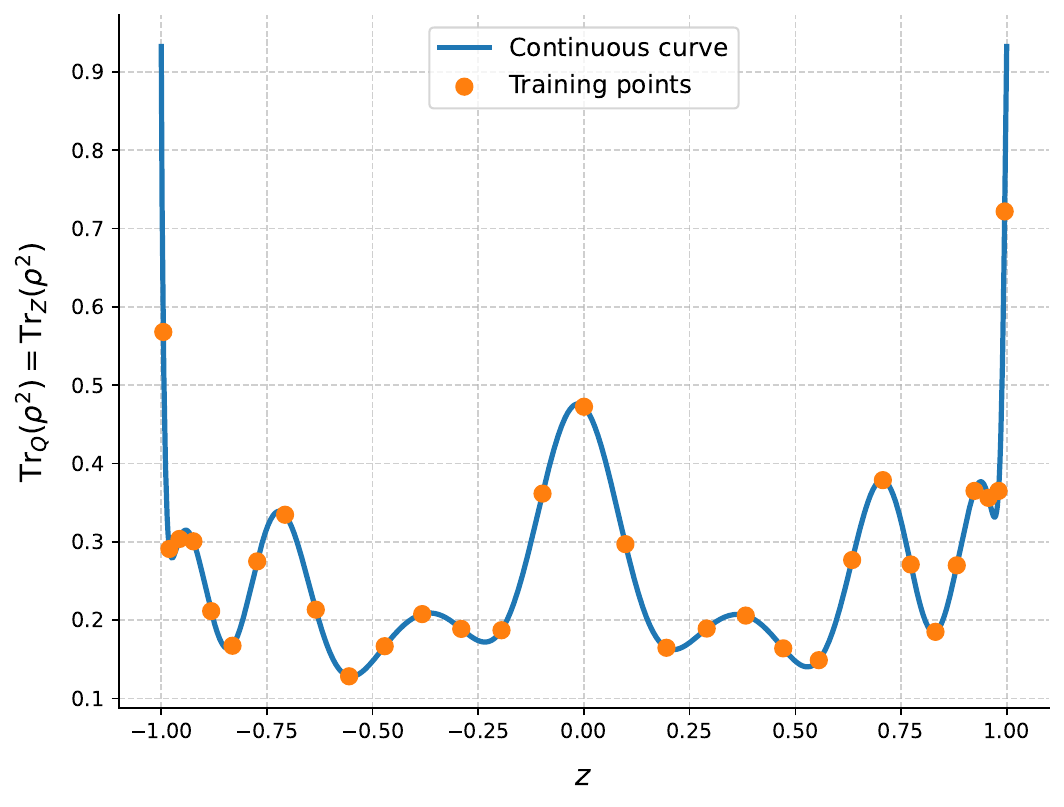}
    \caption{Purity of the quantum register $\mathcal{Z}$ for different $z$ values. Orange points represent the training points in Chebyshev nodes and halfpoints. Blue line represents a continuous curve in the range $[-1,1]$.}
    \label{fig:purity}
\end{figure}

Now we can study how the purity of the register $\mathcal{Z}$ evolves as a function of the momentum fraction $z$. 
As shown in Fig. \ref{fig:purity}, the purity at the Chebyshev nodes (represented by every second orange point) alternates between local minima and maxima. Specifically, the purity reaches its maximum values at the boundaries, $-1$ and $1$, as well as at $0$, indicating minimal entanglement. In contrast, maximal entanglement, corresponding to the lowest purity, occurs at values around $0.5$ and $0.8$.

\subsection{Mutual information in quantum registers}

In quantum information theory, quantum mutual information quantifies the total amount of correlations, both classical and quantum, between two subsystems of a quantum state \cite{nielsen}. For a bipartite system described by a density matrix $\rho_{\mathcal{ZQ}}$, the quantum mutual information between subsystems $\mathcal{Z}$ and $\mathcal{Q}$ is given by:
\begin{equation} 
I(\mathcal{Z}:\mathcal{Q}) = S(\rho_\mathcal{Z}) + S(\rho_\mathcal{Q}) - S(\rho_{\mathcal{ZQ}}) 
\label{eq:qmi}
\end{equation}
where $S(\rho) = -\text{Tr}(\rho \log \rho)$ is the von Neumann entropy. A high mutual information indicates strong correlations between the two subsystems, implying significant entanglement. In contrast, $I(\mathcal{Z}:\mathcal{Q}) = 0$ means the subsystems are completely independent.

An interesting simplification of Eq. \ref{eq:qmi} happens in the case where the matrix $\rho_{\mathcal{ZQ}}$ corresponds to a pure state, as it is our case. In that case $S(\rho_{\mathcal{ZQ}})=0$ and $S(\rho_\mathcal{Z})=S(\rho_\mathcal{Q})$. Hence Eq. \ref{eq:qmi} reads:
\begin{equation} 
I(\mathcal{Z}:\mathcal{Q}) = S(\rho_\mathcal{Z}) + S(\rho_\mathcal{Q}) = 2 S(\rho_\mathcal{Z})= -2 \text{Tr}_\mathcal{Z}(\rho\log_2 \rho).
\label{eq:qmi2}
\end{equation}

\begin{figure}[h] 
\centering 
\includegraphics[width=0.5\linewidth]{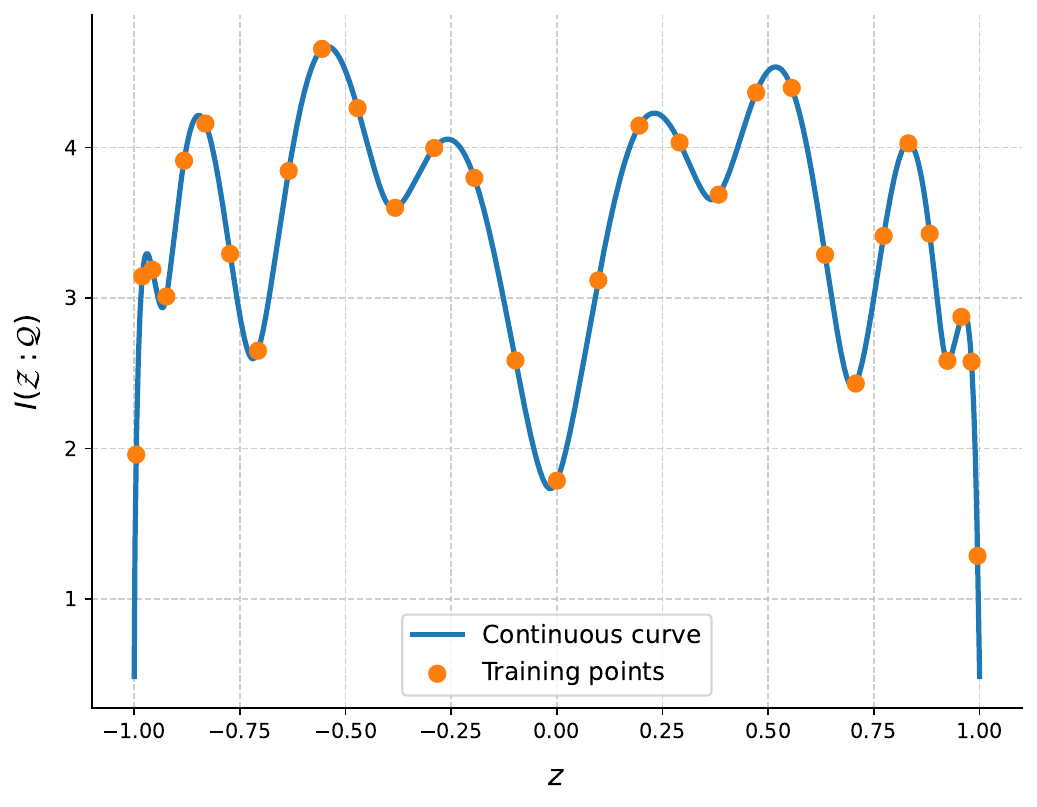} 
\caption{Quantum mutual information between registers $\mathcal{Z}$ and $\mathcal{Q}$ for different $z$  values. Orange points are training points in Chebyshev nodes and halfpoints. Blue line represents a continuous curve over the range $[-1,1]$.} 
\label{fig:mutual} 
\end{figure}

In this section, we quantify the entanglement between the registers encoding $z$ and $Q$, in the training stage of Fig. \ref{fig:2DQCs}(a), using their mutual information. Specifically, we compute the quantum mutual information for the system $\mathcal{Z}$ of $N=5$ qubits, which depends on the variable $z$.
As shown in Fig. \ref{fig:mutual}, the mutual information exhibits alternating local maxima and minima at the Chebyshev nodes. The strongest correlations appear at intermediate values of $z$, where the entanglement is highest, while minimal correlations occur at the boundaries, $-1$ and $1$, as well as at $0$. 

\end{document}